# Membrane Space Telescope:
# Active Surface Control with Radiative Adaptive Optics

S. Rabien*[a], L. Busoni[b], C. Del Vecchio[b], J. Ziegleder[a], S. Esposito[b]
[a]Max Planck Institute for Extraterrestrial Physics, Giessenbachstr. 1, 85748 Garching, Germany;
[b]INAF, Arcetri Astrophysical Observatory, Florence, Italy

## ABSTRACT

Sensitivity and resolution of space telescopes are directly related to the size of the primary mirror. Enabling such future extremely large space telescopes or even arrays of those will require to drastically reduce the areal weight of the mirror system. Utilizing a thin parabolic polymeric membrane as primary mirror offers the prospect of very low weight and the flexible nature of those membranes allows compactly store them upon launch. Upon deployment the structure is unfolded and the mirror shape restored. Being an extremely thin structure, an active shape correction is required. Utilizing a thermal control of the surface via radiative coupling, localized shape changes are imprinted into the membrane telescope. In this paper we present the modelling and experimental test of the radiative adaptive optics. A detailed modeling of the influence function of the radiative shaping onto the membrane mirror has been carried out. Experimentally we have been radiatively actuated the shape of a prototype mirror in closed loop with a wavefront sensor and proven that we can control the mirrors surface figure to a ~15nm RMS precision.

**Keywords:** Space telescope, membrane primary mirror technology, radiative adaptive optics

## 1. INTRODUCTION

Building a space-based telescope and launching and deploying it in orbit is usually an endeavor with high costs and long timescales involved. Enlarging the apertures is highly desirable for astronomical missions or earth observing, for both to increase sensitivity and resolution. As large telescope apertures come with the challenge of weight and volume many attempts to reduce the weight of the primary mirror are made and fold the structure upon launch. Naturally light weight structures are subject to deformation and will require corrective elements, such as deformable mirrors in the optical train. An overview can be found e.g. in [1]. Efforts are in place to space-qualify deformable mirrors (e.g. [3][4][5][6][7]) and design folding structures.

With the goal in mind to significantly enlarge the aperture of a future space telescope, or laying the ground for a space-based interferometer array, we have developed a technique to produce thin and precisely pre-shaped parabolic polymer mirrors [8]. With being a flexible material, those mirrors can be rolled up and stored compactly in a launch vehicle, and then be deployed in space. This thin membrane-based mirror achieves a significant reduction in areal weight of the primary, which in conjunction with the ability to roll the mirrors is a key element to achieve much larger telescope apertures in space at reduced launching cost. As the thin material will not be in the perfect optical shape upon un-rolling, and naturally is subject to shape changes from gravity release and thermal environment, we have developed a surface shaping method based on radiative adaptive optics. With illuminating the mirror by a spatially variable source, we obtain a controlled local surface deformation. This method is contact free, does not require backing structures and adds very low additional weight to the system.

Utilizing a 30cm prototype mirror we have setup set up a closed loop wavefront control by measuring the surface shape with a wavefront sensor and illuminating the surface with a DLP. By closing the loop with a real time control we have obtained a reflected wavefront quality of 35nm RMS. With the expectation that this technology can be scaled up to large telescope diameters, we have the means at hand to achieve diffraction limited performance into the visible spectral range. In this paper we focus on the radiative adaptive optics method. While first giving an overview on the membrane mirror technology and the polymer material we show the outcome of physical modelling of the radiative influence. In the following section we present the laboratory results from the closed loop adaptive control of the prototype mirror.

*srabien@mpe.mpg.de

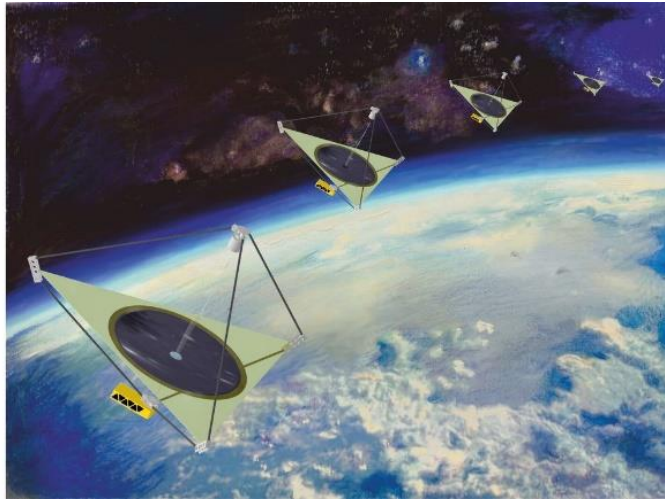

*Figure 1: artists impression of a space telescope array. Maybe a dream nowadays, but such an array of large telescopes could empower us to address science cases requiring extreme sensitivity and resolution. Resolving planets around other stars might be one of the most popular and exciting one. Key technologies for the building blocks are light weight, compactly foldable and optically precise large mirrors.*

## 2. THIN MEMBRANE MIRROR TECHNOLOGY

The mirrors described in this paper are produced by chemical vapor deposition of a polymer onto a rotating liquid mandrel. Well known since a long time, liquids that are contained in a rotating container, with the rotational axis aligned with the gravitational vector, form a surface with parabolic shape. A mathematical description, including surface tension effects can be found in [9]. The effect is used for zenith pointing telescopes [10][11] or spin casting glass mirrors [12].

We have been using this parabolic surface to grow a polymeric membrane, replicating the shape of the liquid. With the focal length being calculated by:

$$f = \frac{g}{2\omega^2}$$

One can see that low f numbers and large telescope diameters lead to significant peripheral speeds, causing winds and distortions. With the CVD process taking place in a vacuum chamber, distortions from winds or dust particles are excluded, resulting in an optical quality liquid surface. The deposition of the parylene polymer is based on the Gorham [13] process. In a custom build reactor, a dimer pre-cursor is evaporated, thermally cracked into monomeric molecules that react to the polymer at room temperature on the surfaces inside of the deposition chamber. To achieve a 200µm thick polymer on the liquid, we typically run the process over several days. When finished, the front surface is overcoated by a reflective metallic layer. Cutting the membrane off the turntable and washing the liquid away leaves us with a free-standing parabolic polymer-based membrane mirror. The mirror used for this paper in the experimental section was produced in our prototype CVD chamber. The size of the vacuum deposition chamber currently allows to grow 30cm diameter membranes. For optical tests the mirror is left to rest on its border, a 2cm wide flat surface that is grown simultaneously with the parabolic surface. Figure 2 shows a photograph illustrating the flexibility of the mirror, that can be rolled and restores its shape when putting it back on its rest.

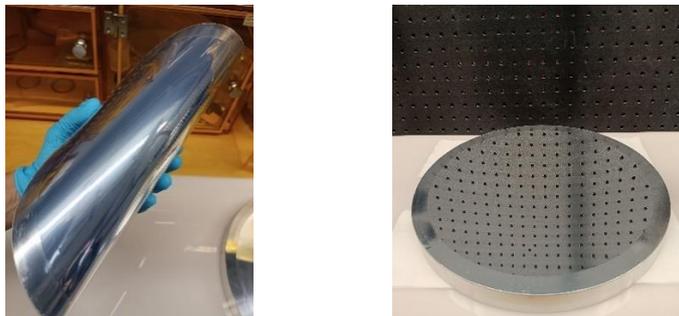

*Figure 2 Left: photograph of the membrane mirror, illustrating the flexibility of the material. Right: photograph while resting on an aluminum ring with its border, reflecting the breadboard in the background.*

## 3. POLYMER MATERIAL PROPERTIES

Poly(para-xylylene) commonly named Parylene as mirror base material owns several desired properties. First of all, the material is formed in a room temperature chemical vapor deposition process at low pressure in a vacuum chamber. This enables the liquid paraboloid shape to stay un-distorted from any environmental effect and keep its mathematically perfect shape. With the monomeric molecules arriving at the surface where the polymer is formed, the layer is intrinsically stress free. No solvent evaporation, nor curing shrinkage or phase change from solidification is involved. Parylene coatings are known for many properties that make them a favorable material for our application. They can be used between cryogenic temperatures up to 200°C dependent on the type, with excellent mechanical properties at cryogenic temperatures. There is basically zero outgassing, an excellent chemical stability, barrier properties to moisture, gases and ionic species. Parylenes have space flight heritage as e.g. as electronics coating for JunoCam and as window onboard the Solar Orbiter and are listed in NASA low outgassing materials database.

Some material properties for Parylene-C that have been used for calculations are listed in Table 1. The values are summarized from [14][15][16][17].

*Table 1: selected properties of Parylene-C used for the calculations and as reference*

| Parylene C-type property | unit | value |
|---|---|---|
| Melting point | °C | 290 |
| Glass transition | °C | 50 |
| Linear thermal expansion | ppm | 35 |
| Thermal conductivity (25°C) | W/mK | 0.084 |
| Thermal capacity | kJ/kgK | 0.712 |
| Continuous service temperature (air) | °C | 80 |
| Continuous service temperature (inert) | °C | 230 |
| | | |
| Youngs modulus | GPa | 2.8 |
| Tensile strength | MPa | 68.9 |
| Yield Strength | MPa | 55.2 |
| Density | g/cm$^3$ | 1.289 |
| Refraction index | | 1.639 |

Parylene is known to be an extremely stable material, basically insoluble in any solvent. It can be used at elevated temperatures up to 230°C in an inert atmosphere. [18] showed that aging on free standing membranes basically does not occur over many years. Since we use the material as an optical substrate long term deformation should be avoided. [19] found that below the glass transition temperature basically no creep occurs in Parylene-C films. We have seen in our membranes stored in the laboratory with the gravity axis perpendicular to the surface no change in optical shape over a year's timescale. While being generally very stable, the C-type Parylene is not recommended to be exposed to UV radiation. While as well other types of the polymer can be used, a space mission of such a mirror would have a metallic coating and will be protected by a sun shield for a variety of reasons.

Optically Parylene is highly transparent over the whole visible spectral range, but shows several absorption lines in the infrared between 3 to 15µm (see e.g. [18]) with prominent lines at 3.3µm, 6.9µm and 9.6µm. As we use radiation to locally heat the polymer for the active optics, a suitable light source may better be tuned to the applied coating, as the absorption lines in the mid infrared do not fit available laser sources optimally.

# 4. RADIATIVE ACTUATION

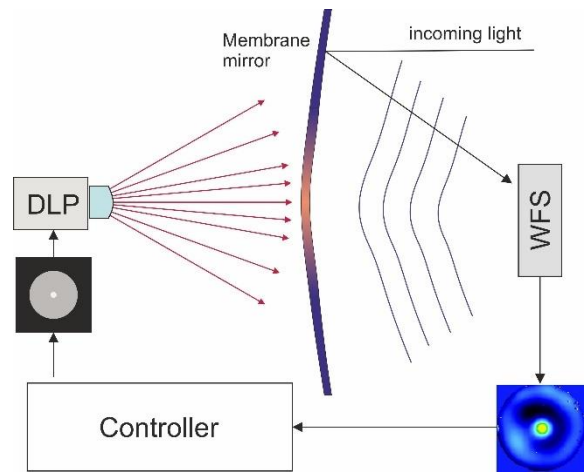

*Figure 3 Sketch of the basic principle of the radiative mirror actuation. The curved membrane mirror is irradiated by a spatially variable light source (DLP) over its entire area. At locations where a local increase of the intensity is commanded, a local increase of mirror curvature is resulting from the materials thermal expansion. The reflected incoming light is directed towards a wavefront sensor (WFS), measuring the shape of the mirrors surface. Feeding the measured wavefront to a control computer one can close the loop with adjusting the commanded light intensity pattern.*

Since the thin and flexible material is subject to deformations from a variety of sources, as would be the gravity release or the un-rolling process, a way to control the surface shape is required. Other membrane attempts have been proposing electrostatic [20][21][22] or piezoelectric methods [23]. We have found that a method, exploiting the thermal expansion of the material can serve this purpose. Figure 3 shows a sketch of the basic principle. A spatially variable light source is irradiating the whole surface. This can be a digital light projector (DLP) or a laser scanning system or similar. With some amount of the light being absorbed by the membrane mirror an increase in temperature is resulting, causing the material to expand. Parylene-C has a CTE of 35ppm, not as high as many polymers but larger than most metals. Since the mirror is a curved surface, any local material expansion causes a local decrease of radius of curvature, and the other way around. Keeping a global radiation bias allows to actuate in both directions: increasing the irradiation adds curvature and decreasing radiation lowers the curvature. With this mechanism we have a mean at hand to locally push and pull the optical shape to our desire. With the membrane mirrors being of an optical quality, good enough to be probed by a wavefront sensor, the shape of the mirror can be measured. Probing the wavefront with e.g. starlight being reflected from the front surface we can feed this into a real time computer that regulates the intensity distribution, being dialed into the DLP system, which closes the loop.

In the experiments described later in this paper we have been utilizing a standard DLP beamer operated at white light and illuminating the front surface of the mirror. With the aluminized surface reflecting 95 percent of the light, only a small amount of the energy is used to heat the material. In a real space telescope setup, one would add a thin gold coating or similar to the front or back surface to act as an absorber. With such a coating a narrow band blue or green light source would be highly absorbed, providing a more efficient coupling, lowering the power requirements and facilitate a suppression in the telescopes scientific light path.

Compared to other methods the radiative actuation of the mirror does not require any second backside structure, as would be needed for electrostatic or magnetic actuation. As such very little additional weight and complexity is needed. The only requirement is a clear path to project the pattern onto the membrane and for any reflected portion to escape to space.

Systems that allow to generate the required programmable light projection are available as scanning systems or with micromirror arrays. Digital micromirror devices are widely used for projection purposes, can own millions of individual mirrors and have seen space qualification efforts [24]. With off-order 1000 elements across the aperture, even on a 20m telescope the individual pixel is still only a 20mm in footprint, small enough to correct for even high frequency surface errors. Manipulating the mirrors shape with the radiative coupling from such a DMD translates it into a deformable mirror with ~800000 actuators.

## 5. MULTI-PHYSICS MODELLING OF THE RADIATIVE INFLUENCE FUNCTION

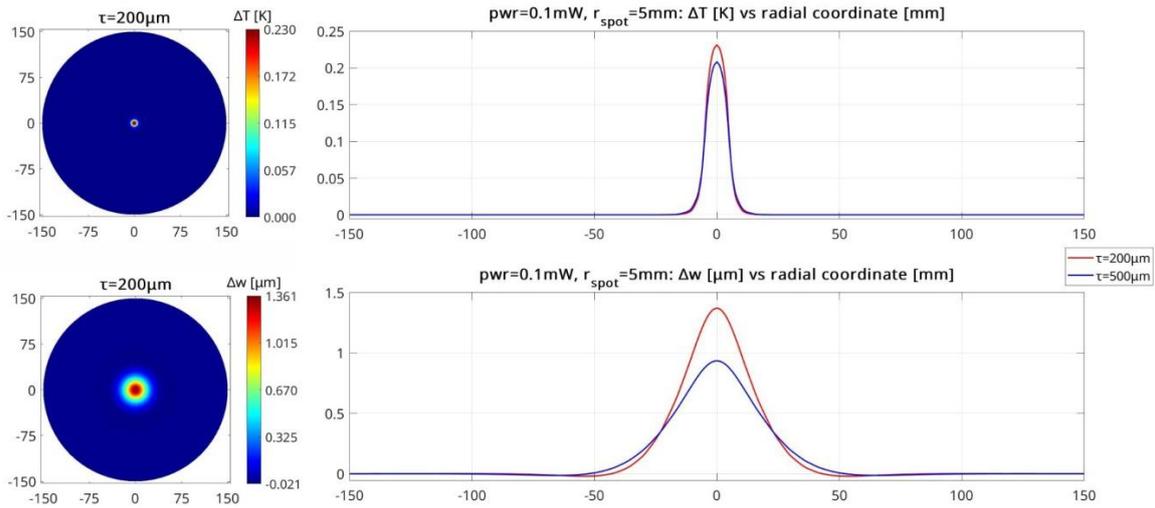

*Figure 4 an example of the calculated response function to the radiation of a 5mm radius single illumination spot. To the top left the temperature distribution within the material for a 0.1mW radiation being absorbed, to the bottom left the material deflection due to the thermal expansion. To the right cuts through the distribution for two thickness cases 200 and 500μm are shown. The FWHM of the 200μm case amounts to 30mm, the 500μm case shows 46mm with a shape of the influence function close to gaussian.*

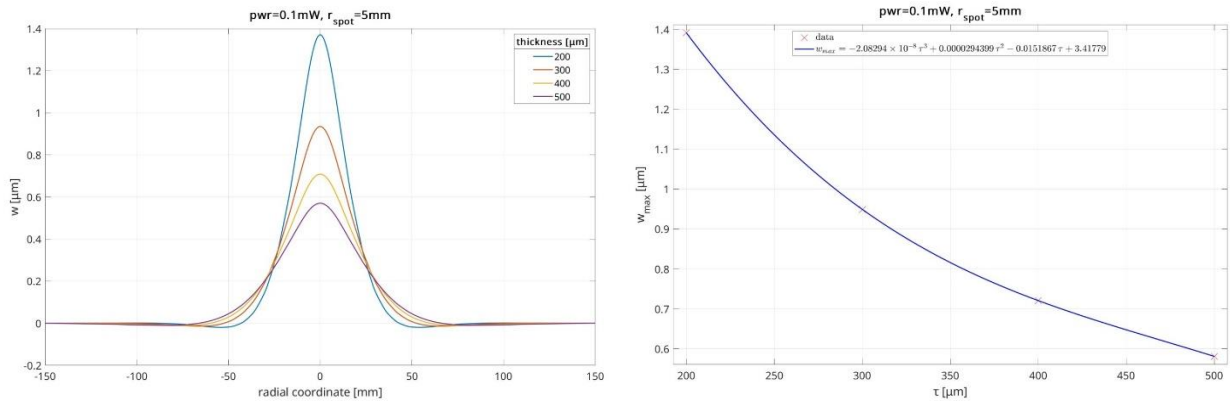

*Figure 5: looking at a variation of membrane thicknesses one can see that at a given radiative power the stroke decreases with increasing thickness with the width of the influence function increasing.*

To understand the response of the membrane mirror to the radiated power we have been setting up a multi-physics simulation of the 30cm prototype mirror. The environmental conditions have been chosen to reflect the laboratory to allow a comparison with the experimental data. Thanks to the simple geometry of the membrane, the simulation has been fully performed within the Comsol Multiphysics program, selecting the optical axis as z. After generating the paraboloid surface and identifying the illuminated spot circular areas, a mesh of ~4500 shell elements, with an appropriate spatial resolution in the spot areas, has been developed. As the thickness of the membrane is much lower than its diameter, the structural response is very well defined with this approach. Inhibiting all the translational degrees of freedom of the central node as well as its x and y rotational degrees of freedom and restraining the azimuthal degree of freedom of a node on the outer circumference allows to get an isostatic system, representative of the operational response.

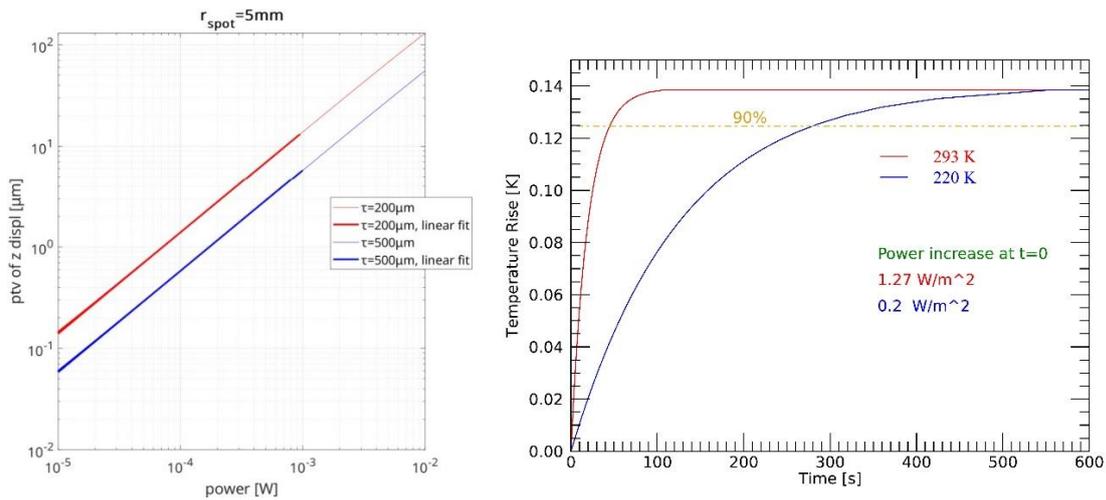

*Figure 6 Left: peak displacement in dependence of the absorbed radiation power. Over a wide range the relation of the peak stroke to the incident power is linear. Right: temporal step response of the temperature when applying a radiation increase. Two cases are shown: the mirror at room temperature and at 220K. With the lower radiation coupling at lower temperatures the response time is increasing as expected.*

As we regard here the laboratory case a radiative coupling of the above defined structure with the ambient is done at 293 K. This non-linear solution of the coupled equations of the resulting physics, with a total number of ~64500 degrees of freedom, has been performed as a function of the following three parameters: the input power, the radius of the spot circle, and the thickness of the membrane. The computation has been performed with selecting a power range spanning from 0.1 to 1 mW, a spot radius ranging from 5 to 30 mm, and a 200 and 500 µm as thickness. These two thickness values are selected since the test membrane is 200µm thick, but future large membranes might be grown thicker. Figure 4 shows the calculated temperature distribution and the displacement influence function for a 10mm spot radius. In Figure 6 the calculated peak displacement versus the absorbed radiation power is displayed. Over a large range the absorbed power is linearly related to the achieved stroke. Figure 7 summarizes the displacement calculations for the 200µm and the 500µm case. In summary we find a linear relation between power and stroke up to mW level of irradiation. Single spot displacements do reach ten microns at mW level of radiation power. The temporal response of the temperature increase to a step of radiation increase is shown in Figure 6 to the right. In this purely radiative coupled calculation one can deduct mainly that at lower temperatures the response time will increase. The room temperature calculation here can not directly be compared to the experimental data shown in Figure 11, since the emissivity of the coatings is not included, and the experiments are carried out under laboratory conditions with air thermal conductance and convection.

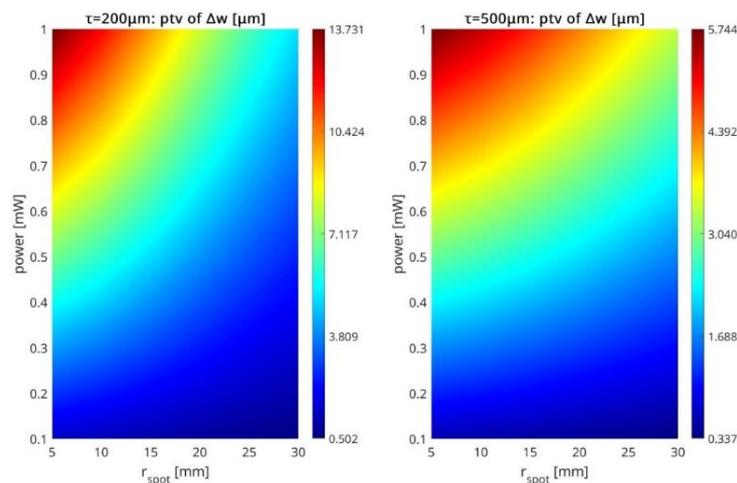

*Figure 7. color coded plot for the achievable stroke for an input power between 0.1mW and 1mW and a spot radius between 5mm and 30 mm. To the left the 200µm thick membrane is shown, to the right the 500µm case.*

# 6. OPTICAL TEST SETUP

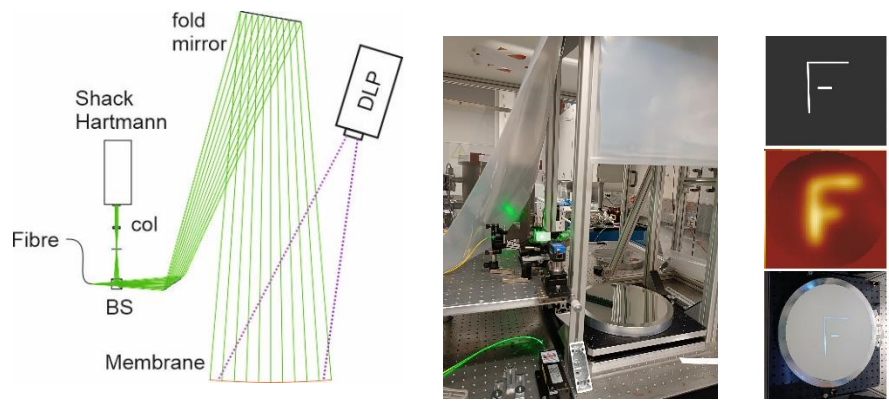

*Figure 8: Optical setup for the surface shape measurement and manipulation. A single mode fibre output is located at the radius location of the membrane curvature. The fibre itself is fed by a 635nm diode laser with adjustable power. The beam is reflected back by the membrane mirror, and split off by a cube beam splitter (BS) towards the wavefront sensor. A collimating doublet lens (col) re-images the membrane mirror onto the lenslet array of the Shack Hartmann sensor. A flat fold mirror in between is only implemented for the limited headroom above the optical bench. The DLP is mounted off-axis with the light projection onto the front surface of the membrane mirror. To the right: top the file dialed into the beamer, mid: the phase reaction, bottom: the letter seen in scattered light on the mirror.*

To measure the membrane mirror and the radiative actuation we have set up a 'test tower' with the membrane resting on bottom. The membrane mirror itself just rests on a milled aluminum ring, supporting its outer border. As a light source a single mode fibre is fed by a 635nm laser and the output located at the radius of curvature of the mirror, 1800mm away from the surface. The light reflected by the mirror is split off before the focus by a cube beam splitter and directed towards the wavefront sensor. A lens collimates the beam after the focus and re-images the mirror onto the Shack-Hartmann lenslet array. The commercial Shack Hartmann sensor from Optocraft provides a high dynamic range and many sub-apertures for the measurement. The DLP, a commercial image projector, is located approximately 1m above and aside the membrane, projecting downwards. The beamer is enclosed in a box with window and the generated heat ventilated away. The off-axis location of the beamer is not optimal, as it generates the need to adjust the key stone setting and results in a slight uneven illumination power over the membrane. Figure 8 shows a sketch from the setup, a photograph and an example file of the phase reaction to the DLP illumination.

In Figure 9 the recorded SH spot pattern is shown. Typically, 85 sub-apertures over the center of the membrane mirror are measurable, spanning 215mm in diameter. The borders are affected by the deformation under gravity, border tension effects from the grows and the milling precision of the aluminum rest, leaving an outer ring of ~5cm with too high gradients for the sensor. For the active optics test described in the following chapter mostly a measurement diameter of 77 sub-apertures, spanning 195mm was used.

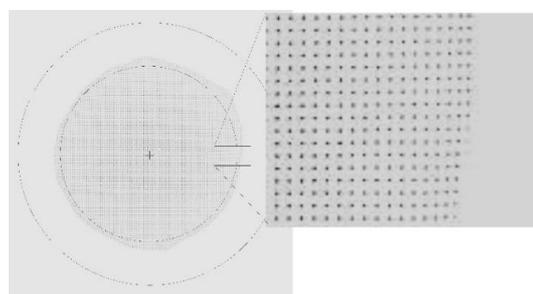

*Figure 9: Shack-Hartmann spot pattern as seen by the sensor. 85 sub-apertures are measurable across the mirror. The outer 5cm show too high gradients to be seen by the sensor, which can be understood as being caused by the deformation of the membrane under gravity. The inner circle in the left plot denotes a 200mm diameter, the outer one marks 300mm.*

# 7. ADAPTIVE SURFACE SHAPE CONTROL

We have set up the system as described above to test the ability of radiative adaptive optics to remove wavefront aberrations from the membrane mirror system. As such, our prototype membrane mirror is just placed on its aluminum rest without making a big attempt to compensate for the surface deviations induced by the border rest. This usually induces global deformations of some 10 to20 µm peak to valley that we intend to correct with the radiative adaptive optics.

**Geometry and power calibration**

In order to calibrate the geometry of the setup we measure the magnification of the beamer pixel on the mirror surface by projecting a circle of 540 pixel radius to fully overlap over a rigid membrane cover having a circular pattern of 150mm radius. As such, the measured magnification amounts to 3.6pixel/mm.

In a similar way we estimate the projected Shack Hartmann wavefront sensor sub-aperture size on the mirror by thermally inducing two indentations at a known physical distance of 111mm that are seen spaced by 43.6 subapertures in the wavefront map, corresponding to a WFS sampling of 0.394 subap/mm.

The largest area that can be monitored by the WFS before saturation is about ~77 subaps, corresponding to about 195mm diameter on the membrane surface. The DLP intensity has been calibrated with focusing a projected circle on a power meter as shown in Figure 10. The DLP is set up with a Gamma correction of 1.8 (the intensity is proportional to the bitmap level to the power of 1.8). At full intensity (grey level = 255) the DLP delivers about 1.9W over an area of 1038 pixel diameter, corresponding to ~2.3µW/px incident, or about ~0.1µW/pix absorbed by the membrane. As such, a typical 20px radius circle as used in the following, deposits ~0.12mW on the membrane, assuming a reflectivity of the aluminum of 0.95 over the RGB output.

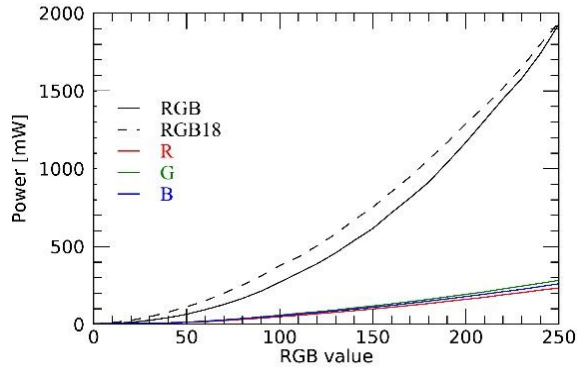

*Figure 10: power output of the beamer in R,G,B and white light against the commanded 8-Bit value. In the experiments described here we have used the RGB18 setting, delivering enough power, and being closer to a linear behavior. More tailored light sources for this purpose will easily be able to deliver the wished response and higher intensity resolution.*

**Influence Function**

A membrane deformation is induced by illuminating with the beamer a circular area: a spot of a few millimeters' diameter collects enough energy to create a wavefront indentation of about 2-3 micron with the DLP set to full power. Figure 11 shows a single spot influence function measurement. With the FWHM arriving at 30mm and the peak displacement of about 3µm the measurement well agrees with the FEA calculations as shown in section 5. The temporal response of the deformation is reported in Figure 11 to the right. The response is well modeled with a function like $z(t) = A * [1 - \exp(-t/\tau)]$ with a 0-90% rise time of about 37 seconds. As discussed above, we expect the temporal response to be dependent on the radiative coupling to the environment. At lower temperature of the system, the response will get generally slower.

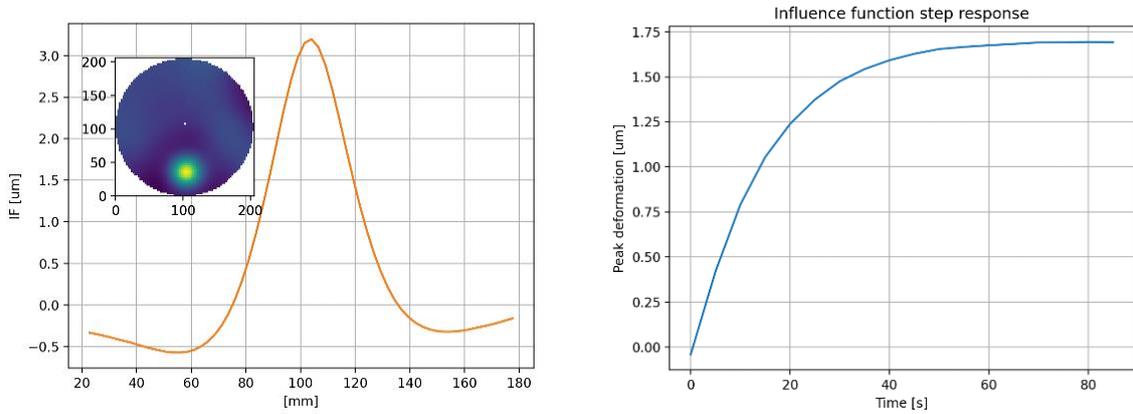

*Figure 11 Left: Cut through the influence function (wavefront map in the inset) corresponding to a 5.5mm radius illuminated circle at full beamer intensity (corresponding to about 0.12 mW absorbed power). The FWHM of about 30mm and a wavefront deformation of about 3um are in good agreement with the results of FEA considering errors from the power distribution of the DLP, being not completely homogeneously over the membrane. Right: measured temporal step response function of this illumination.*

The response to the beamer grey levels is shown in Figure 12. The peak of the deformation is proportional to the grey-level to the power 1.8, as expected from a membrane showing a deformation proportional to the absorbed power and considering the DLP gamma correction set to 1.8. The influence function peak is also proportional to the illuminated area, in this range of circles diameter ranging from 5.5 to 22mm.

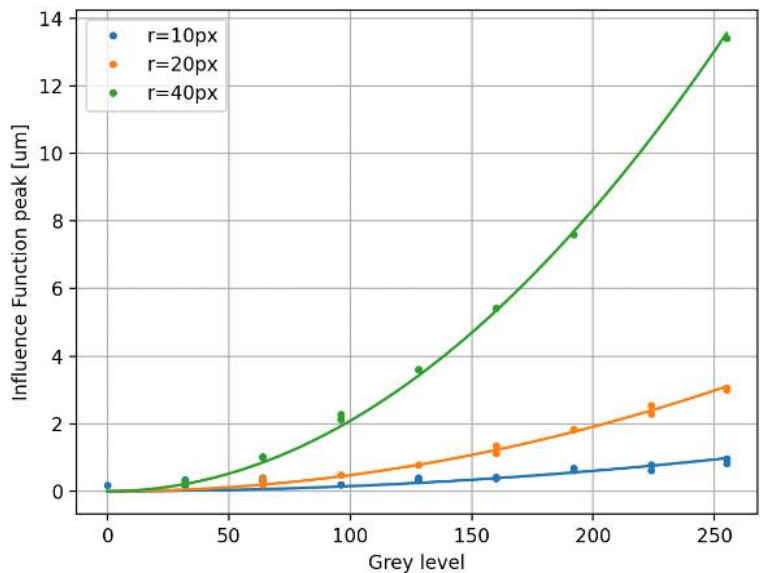

*Figure 12 Influence function amplitude as a function of DLP grey level for 3 circles of different radius. The membrane deformation follows quite accurately a power law 1.8, as expected for a DLP with a gamma correction 1.8 and a membrane with a deformation linearly depending on the absorbed power.*

**Mirror shape at rest**

As we apply a bias radiation to allow a bi-directional actuation we have evaluated the influence of a constant illumination on the reflected wavefront. The wavefront of the membrane without illumination and when a bias illumination of rgb=128 is applied, is shown in Figure 13.

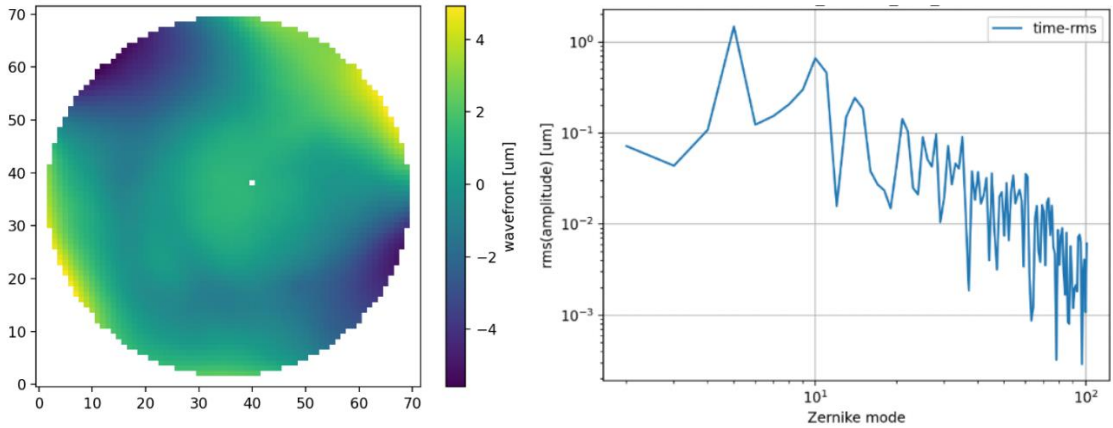

*Figure 13 Typical wavefront map with the membrane at rest (left) and modal decomposition on 100 Zernike polynomial modal basis.*

The wavefront error is 1.8-2 µm rms wavefront, about 10-14 µm ptv, dominated by the first 20 Zernike modes whose amplitudes range in 100-1000nm. After proper thermal and acoustic insulation of the setup, the wavefront stability is quite good, about 20nm with the bias illumination and a few nm when the DLP is switched off (see Figure 14). The increase in wavefront change upon illumination can be understood from the energy deposited in the optical system, causing phase variation from variable heating of the air environment and feedback to the membrane. Such dynamical variations are expected to be only present in the laboratory environment, while it is much quieter in a shielded dark space environment.

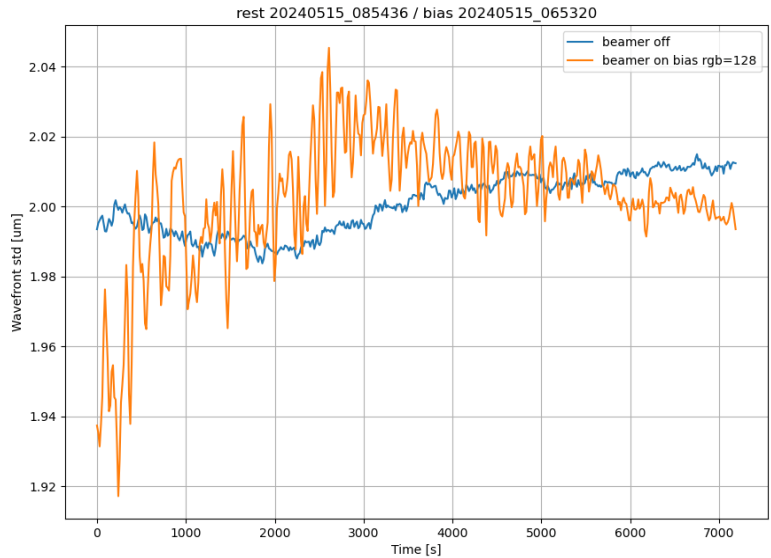

*Figure 14 Wavefront stability. Membrane under constant bias illumination (orange) and membrane at rest, with DLP off (blue).*

**Mirror figure correction**

The correction of the membrane mirrors is obtained with a zonal control loop between the Shack Hartmann wavefront sensor and the DLP beamer. The geometrical transformation to map the beamer reference system over the WFS subapertures is measured by illuminating the membrane with a 3x3 spot grid of 11mm spot diameter and a spacing of 200 beamer's pixels, corresponding to about 55mm on the mirror surface.

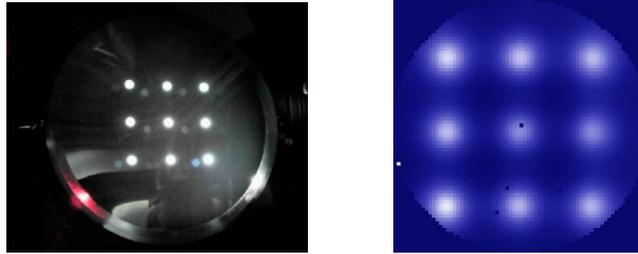

*Figure 15 Left: Illumination pattern on the membrane during geometry calibration as seen by a monitoring webcam. Right: the corresponding wavefront deformation measured by the wavefront metrology.*

The location of the indentations maxima in the WFS reference system is used to solve a first order distortion system. The inverse geometrical transformation is then used to convert the wavefront measured by the WFS onto the beamer reference system. Since the membrane area monitored by the wavefront sensor is smaller than the full aperture, we have to address the issue of controlling the shape of the unseen membrane: we implemented a slaving procedure to extend the beamer commands outside the measured area. The procedure projects the commands on about 50 modes over a circle 20% larger and applies a smoothing filter to avoid steep gradients. An example of a beamer command map is shown in Figure 16.

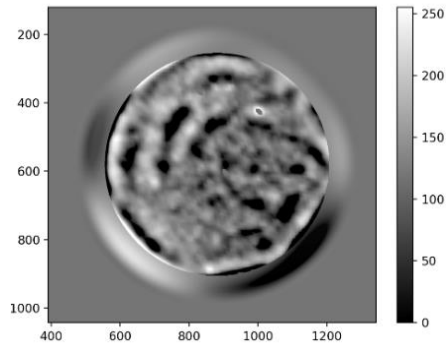

*Figure 16 Command map to the beamer in one of the closed loop approaches. The labels are the coordinates in pixel on the beamer.*

Since the irradiation creates an indentation in the membrane (very much like in electrostatic MEMS mirrors where the actuators can only pull the membrane) we apply a global bias corresponding to about 120 grey levels in such a way that deformations can be applied in both directions when increasing or decreasing the illumination.

A zonal control loop has been implemented with a framerate of 15s, and an integral control with a 3dB bandwidth of ~1mHz. Figure 17 shows an example of a closed loop convergence. With the step response of order 40s and an adapted small gain being applied, the convergence takes a while. Running the loop over typically two hours leads to a convergence of the wavefront deviation. While we do see dynamic distortions from the laboratory environment sometimes dominating the control, an adequate shielding helped in that context. In a better controlled environment and in vacuum space conditions we would expect an even better surface control. With the initial membrane deformation leading to typically about 1.8µm rms wavefront distortion it can be corrected down to 30nm rms wavefront deformation with the radiative adaptive optics. This converts to a ~15nm rms surface quality of the polymer membrane mirror with this adaptive optics correction in place.

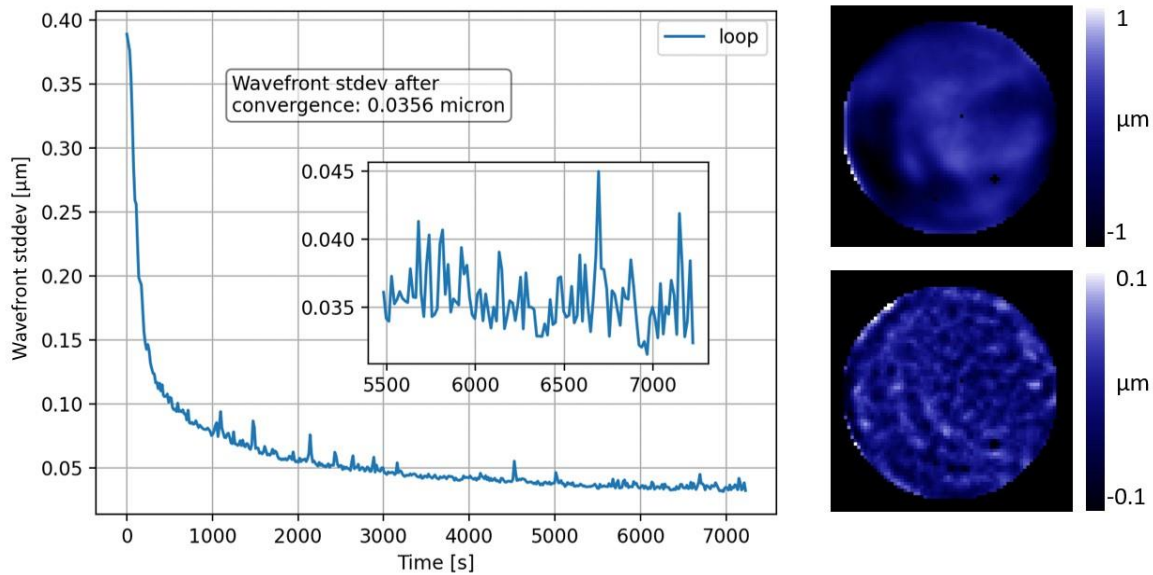

*Figure 17: Measured wavefront standard deviation during loop convergence. The typical starting RMS of the wavefront amounts to 1.8 μm. After a 2h loop convergence time we arrive at 0.03μm wavefront distortion. To the right the upper panel shows the wavefront map before, the lower one after convergence.*

## 8. CONCLUSION

We have explored how a thin parabolic polymer mirror for a large space telescope can be controlled in shape by radiative adaptive optics. By irradiating the mirror with a spatially controllable light source one can correct the shape after deployment in space. We have carried out finite element calculations to understand the influence function of the radiative actuation and find those in good agreement with the experiment. With a prototype radiative adaptive membrane mirror system, we could correct initial shape deviations in the micron range down to a ~15nm surface quality by actuating with the radiation from a spatially controllable light source in closed loop. While there is still a lot to explore, the result presented in this paper makes us confident that large membrane space telescopes, operating diffraction limited at visible wavelength, can be realized.